\documentstyle[12pt]{article}
\begin{document}
\bibliographystyle{unsrt}
%
\def\boxit#1{\vbox{\hrule\hbox{\vrule\kern3pt
         \vbox{\kern3pt#1\kern3pt}\kern3pt\vrule}\hrule}}
\setbox1=\vbox{\hsize 33pc  \centerline{ \bf Institut f\"ur Theoretische 
Physik der Universit\"at Regensburg} } 
$$\boxit{\boxit{\box1}}$$
\vskip0.8truecm
\hrule width 17. truecm  \vskip1pt \hrule width 17. truecm height1pt
\vskip3pt
\noindent
May 1994\hfill{TPR-94-13}
\vskip3pt
\hrule width 17. truecm height1pt \vskip1pt \hrule width 17. truecm 
\vskip 1.5cm 
\begin{center}
\huge
The nucleon as a relativistic\\
quark-diquark bound state \\
with an exchange potential \ ${^*)}$. \\
\vspace{1cm}
\Large
H. Meyer \\
\vspace{1cm}
Institut f\"ur Theoretische Physik, Universit\"at Regensburg \\
D-93040 Regensburg, Germany \\
\vspace{1cm}
May 25, 1992 \\
\normalsize
\end{center}

\begin{abstract}
Treating the quark and diquark as elementary particles, the Bethe-Salpeter
equation for the nucleon is solved numerically.
The dependence of the mass on the diquark mass and on the coupling constants is
investigated. The resulting relativistic
quark-diquark-nucleon vertex is presented and discussed. 
\end{abstract}

\vspace{7cm}

${^*)}$ Work supported in part by BMFT grant 06 OR 735 

\newpage
\input epsf
\newcommand{\be}{\begin{equation}}
\newcommand{\ee}{\end{equation}}
\newcommand{\bea}{\begin{eqnarray}}
\newcommand{\eea}{\end{eqnarray}}

\section{Introduction}
The quark-diquark picture has been successful in describing
a variety of nucleon properties \cite{Ans93}. 
It has been used to relate such different phenomena as the $\Delta$-$N$ mass 
difference \cite{GRG75}, the neutron formfactor \cite{IKS81}
and the ratios of structure functions in
deep inelastic scattering \cite{Clo79,MM91}.
In most applications one uses purely phenomenological  
quark-diquark wave functions.
In order to actually calculate the dynamics of the nucleon, some authors have
solved the
Bethe-Salpeter equation for the quark-diquark system,
with a local (quenched or static) quark-diquark interaction
\cite{BAR92}. The integral
equation is then separable and can be solved analytically. The exchanged 
quark does not propagate in this approximation.

In this letter, the quark-diquark-nucleon (qdN) vertex is calculated, using 
the Bethe-Salpeter equation (BSE)
for the quark-diquark system, with a propagating quark exchange potential. The
properly normalized qdN vertex is shown to lead to the correct nucleon
charges. Similar calculations, without inclusion of 
vector diquarks, have been presented in \cite{BCP89}. 

The three-quark problem with a quark exchange potential has recently been solved by Ishii,
Bentz and Yazaki \cite{IBY93}.
However, they did not handle the problem of normalization and did not present
results for the vertices (see our figure (\ref{fig:vertex})), which are important e. g. for 
the calculation of formfactors and structure
 functions \cite{MST94}.
\section{The model}
We start from the following model Lagrangian:
\bea
	L= &i& \bar{\psi} \gamma^\mu \partial_\mu \psi - m_q \bar{\psi} \psi +
	(\partial _\mu \chi)^\dagger (\partial ^\mu \chi) - 
	m_s^2 \chi^\dagger \chi - {1\over 4} F_{\mu \nu} F^{\mu \nu} + 
	{1\over 2} m_v^2 \chi_\mu \chi^\mu \\
 + &i& g_s \psi^T C^{-1} \gamma ^5 \tau^2 \psi \chi^* - 
		i g_s^* \bar{\psi} \gamma ^5 C \tau^2 \bar{\psi}^T \chi + 
		g_v \psi^T C^{-1} \gamma ^\mu \tau^i \psi \chi_\mu^{i*} - 
	       g_v^* \bar{\psi} \gamma ^\mu C \tau^i \bar{\psi}^T \chi_\mu^i ,
		\nonumber
\eea
where $\psi$ is the quark field and $m_q$ is the (constituent) quark mass.
The scalar and vector diquark fields are denoted by 
$\chi$ and $\chi^\mu$. The vector diquark field-strength tensor is denoted by
$F^{\mu \nu} \equiv \partial ^\mu \chi^\nu -\partial ^\nu \chi^\mu$.
The isospin index of the vector diquark is $i=-,0,+$, and
$\tau^-=(1-\tau^3)/2$, $\tau^0=1/\sqrt{2}$ and
$\tau^+=(1+\tau^3)/2$ are the corresponding combinations of isospin
$SU(2)$ matrices for the quarks.
The strengths of the quark couplings to the scalar 
and axial vector diquarks 
are $g_s$ and $g_v$, respectively. 
The charge 
conjugation operator is $C=i\gamma^2 \gamma^0$.

\begin{figure}
\epsfxsize=14cm
\centerline{
		\epsfbox{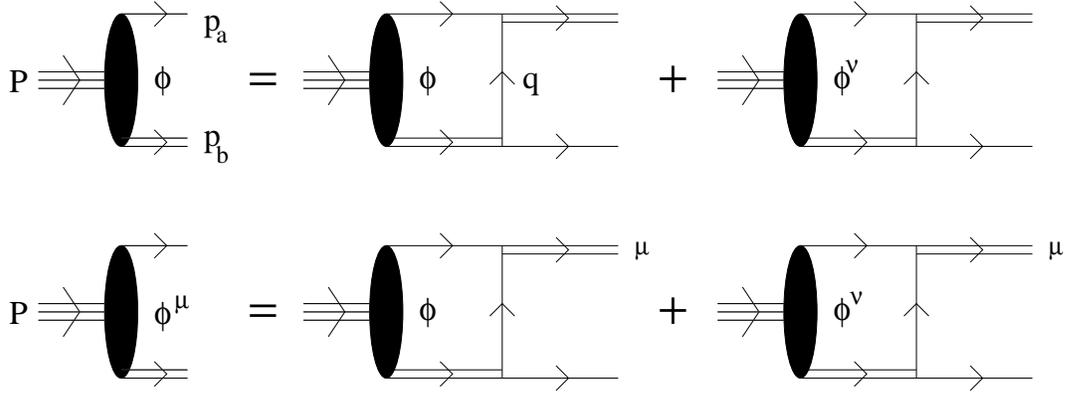}}
	\caption{{\sf The mixing of scalar $\phi$ 
	and vector $\phi^\mu$ quark-diquark-nucleon
	vertices in
	the Bethe Salpeter equation.}}
	\label{fig:BSE}
\end{figure}

The Bethe-Salpeter amplitude of the quark-diquark system
 has scalar and vector components $\phi (p,P)$
and $\phi ^\mu (p,P)$. Their mixing in the BSE is shown in fig. (\ref{fig:BSE}).
The total and relative momenta in terms of the quark momentum $p_a$ and
the diquark momentum $p_b$ are:
\bea
	p &=& \eta_b p_a - \eta_a p_b \\
	P &=& p_a + p_b ,
\eea
where $\eta_a=m_a/(m_a+m_b)$ and $\eta_b=m_b/(m_a+m_b)$ with the quark mass
$m_a \equiv m_q$, and $m_b$ is either one of the diquark masses
$m_s$ or $m_v$.
We redefine the Bethe-Salpeter amplitudes as follows:
\bea
	\tilde{\phi} (p) &=& \left[ iS(p_a) iS(p_b) \right] ^{-1} \phi (p) 
	\label{eq:phi} \\
	\tilde{\phi}_\mu (p) &=& \left[ iS(p_a) iS^{\mu \nu} (p_b) 
				 \right]^{-1} \phi^\nu (p) , \label{eq:phimu}
\eea
thereby removing the external quark and diquark propagators:
\bea
	S(p_a)&=&{/\hspace{-2mm} p_a + m_a \over p_a^2 -m_a^2}, \\
	S(p_b)&=&{1 \over p_b^2 - m_s^2}, \\
	S^{\mu \nu} (p_b)&=&{-g^{\mu \nu} +
	     p_b^\mu p_b^\nu \over p_b^2 - m_v^2}.
\eea
The amplitudes (\ref{eq:phi},\ref{eq:phimu}) satisfy the following equations:
\bea
	\tilde{\phi} (p) =&& 12 i \vert g_s \vert ^2 
	\int {d^4 p^\prime \over (2\pi)^4}
	S (q)
	S (p_a^\prime ) S (p_b^\prime) 
	\tilde{\phi}(p^\prime)\\
	&&-18 i g_s g_v^*
	\int {d^4 p^\prime \over (2\pi)^4}
	\gamma^\mu \gamma^5
	S (q)
	S (p_a^\prime ) S_{\mu \nu} (p_b^\prime) 
	\tilde{\phi}^\nu (p^\prime), \\
	\tilde{\phi}^\mu (p) =&& 12 i g_s^* g_v 
	\int {d^4 p^\prime \over (2\pi)^4}
	S(q)	
	\gamma^5 \gamma^\mu
	S (p_a^\prime ) S (p_b^\prime) 
	\tilde{\phi}(p^\prime)\\
	&&+6 i \vert g_v \vert^2
	\int {d^4 p^\prime \over (2\pi)^4}
	\gamma^\lambda \gamma^5 
	S(q)	
	\gamma^5 \gamma^\mu
	S (p_a^\prime ) S_{\lambda \rho} (p_b^\prime) 
	 \tilde{\phi}^\rho (p^\prime).
	\label{eq:BSbig}
\eea
Here, $\tilde{\phi}^\mu$ is defined such that it involves
the isospin $"+"$ component
of the diquark in the proton
(the component with two up quarks in it). The vector-diquarks in the proton
can have isospin
$"+"$ (up-up) or $"0"$ (up-down) components. 
The corresponding vertices are related
through isospin symmetry: 
$\phi_\mu^{p+}=-\sqrt{2}\phi_\mu^{p0} \equiv \phi_\mu$.
The isospin $"-"$ and $"0"$ vertices in the neutron satisfy:
$\phi_\mu^{n-}=-\sqrt{2}\phi_\mu^{n0} = -\phi_\mu$.
The momentum of the exchanged
quark is denoted by $q$.
\section{Structure of the solution}
The equations (\ref{eq:BSbig}) describe an on-shell nucleon, with its spin 
${1\over 2}$. 
Therefore
the vertex functions $\tilde{\phi}$ and $\tilde{\phi} ^\mu$ can be written as:
\bea
	\tilde{\phi} &=& \Psi U(P,S) \\
	\tilde{\phi}^\mu &=& \Psi^\mu U(P,S),
\eea
where $U(P,S)$ is the nucleon spinor and $\Psi$ and $\Psi^\mu$ are
Dirac matrices. Multiplying the equations (\ref{eq:BSbig})
by $\bar{U}(P,S)$ and summing over the nucleon spin orientation
$S$, it is seen that
only the projections:
\bea
	\chi &\equiv& \Psi \Lambda^+ \\
	\chi^\mu &\equiv& \Psi^\mu \Lambda^+.
\eea
occur, where $\Lambda^+ = (/\hspace{-3mm}P + M)/(2 M)$ is the positive energy
projection operator.
The on-shell vertex functions $\chi$ and $\chi^\mu$
are Dirac operators that satisfy:
\bea
	\chi &=& \chi \Lambda^+ \\
	\chi^\mu &=& \chi^\mu \Lambda^+ .
\eea
They depend on the total 4-momentum $P$ of the nucleon and the relative 
4-momentum $p$ of quark and diquark. Introducing the transverse relative
momentum $p_T^\mu = p^\mu - P^\mu {P\cdot p \over M^2}$, one can make
the following decomposition into scalar functions:
\be
	\chi = S_1 \left( 1 + {/\hspace{-3mm}P \over M} \right) +
	      S_2 \left( /\hspace{-2.5mm}p_T - 
	      {i\over M} p\cdot \sigma \cdot P \right)
	      \label{eq:Sdec}
\ee
for the scalar diquarks 
( $p\cdot \sigma \cdot P = p^\mu \sigma_{\mu \nu} P^\nu$ ),
 and:
\bea
	\chi^\mu = && V_{1} \gamma^5 p_T^\mu
	\left( 1 + {/\hspace{-3mm}P \over M} \right) +
	V_{2} \gamma^5 P^\mu \left( 1 + {/\hspace{-3mm}P \over M} \right)
	 + \label{eq:Vdec}\\
	&& {V_{3} \over M^2} \left( M\epsilon ^{\mu \nu \rho \sigma} \gamma_\nu
	p_\rho P_\sigma + M^2 \gamma^5 \sigma^{\mu \nu} p_{T\nu} +
	\gamma^5 p\cdot \sigma \cdot P P^\mu \right) + \nonumber \\
	&& V_{4} \gamma^5 \left( \gamma_T^\mu -
	 {i\over M} \sigma^{\mu \nu}P_\nu \right)
	+V_{5} \gamma^5 p_T^\mu \left( /\hspace{-2.5mm}p_T -
	 {i\over M} p\cdot \sigma \cdot P \right) + \nonumber \\
	&& V_{6} \gamma^5 P^\mu \left( /\hspace{-2.5mm}p_T -
	 {i\over M} p\cdot \sigma \cdot P \right)
	\nonumber 	
\eea
for the vector diquarks. The eight scalar functions 
$S_{1,2}$ and $V_{1\cdots 6}$, 
depend only on the
scalars $p\cdot p$, $p\cdot P$ and $P\cdot P =M^2$ (which is fixed).
In the static limit, the vertices do not depend on the relative momentum.
Therefore only the functions $S_1$, $V_2$ and $V_4$ survive in this limit.
In the non-relativistic limit, these functions should dominate, since the
remaining functions are multiplied by positive powers of the relative 3-momentum.
\section{Bethe-Salpeter norm and nucleon charge}
From the inhomogeneous BSE, one can derive a normalization condition for
the homogeneous BSE \cite{Nak69}. Defining the quark and diquark parts of
the norm in the S(calar) and V(ector) channels:
\bea
	N_{qS} &\equiv& i \int {d^4 p \over (2\pi )^4}
	{1 \over 2} tr \bar{\chi} S(p_a) {/ \hspace{-3mm} P \over M}
	S(p_a) S(p_b) \chi \\
	N_{dS} &\equiv& i \int {d^4 p \over (2\pi )^4}
	{1 \over 2} tr \bar{\chi} S(p_a) {2 P\cdot p_b \over M}
	S^2(p_b) \chi \\
	N_{qV} &\equiv& i \int {d^4 p \over (2\pi )^4}
	{1 \over 2} tr \bar{\chi}^\alpha S(p_a) {/ \hspace{-3mm} P \over M}
	S(p_a) S(p_b)_{\alpha \beta} \chi^\beta \\
	N_{dV} &\equiv& i \int {d^4 p \over (2\pi )^4}
	{1 \over 2} tr \bar{\chi}^\alpha S_{\alpha \mu}(p_b) 
	{1 \over M} ( -2 P\cdot p_b g^{\mu \nu} + p_b^\mu P^\nu +
	P^\mu p_b^\nu ) S_{\nu \beta}(p_b) \chi^\beta ,
\eea
the BS-norm of the nucleon can be expressed as a linear combination of
these one-loop integrals:
\be
	\eta_a^s N_{qS} + \eta_b^s N_{dS} + {3\over 2}
	( \eta_a^v N_{qV} + \eta_b^v N_{dV} )=1.
	\label{eq:norm}
\ee
Calculating the electric formfactors of proton and neutron at zero momentum
transfer, one finds for the charges of proton and neutron:
\bea
	Q_p&=&{2\over 3} N_{qS} +{1\over 3} N_{dS} + {3\over 2} N_{dV} ,
		\label{eq:Qp} \\
	Q_n&=&-{1\over 3} N_{qS} + {1\over 2} N_{qV} +
	       {1\over 3} N_{dS} - {1\over 2} N_{dV}.
		\label{eq:Qn}
\eea
If the vertex functions (\ref{eq:Sdec}, \ref{eq:Vdec}) are normalized according to
eq. (\ref{eq:norm}), then these charges should be $1$ and $0$. 
\section{Results and discussion}
\begin{figure}[h,t]
\centerline{	\epsfxsize=9cm
		\epsfbox{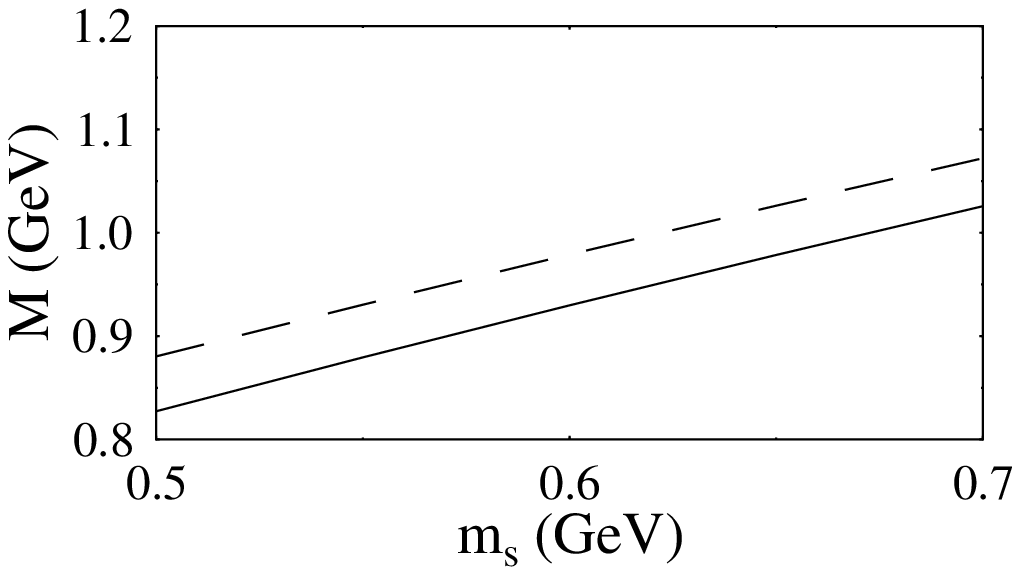}
		\epsfxsize=9cm
		\epsfbox{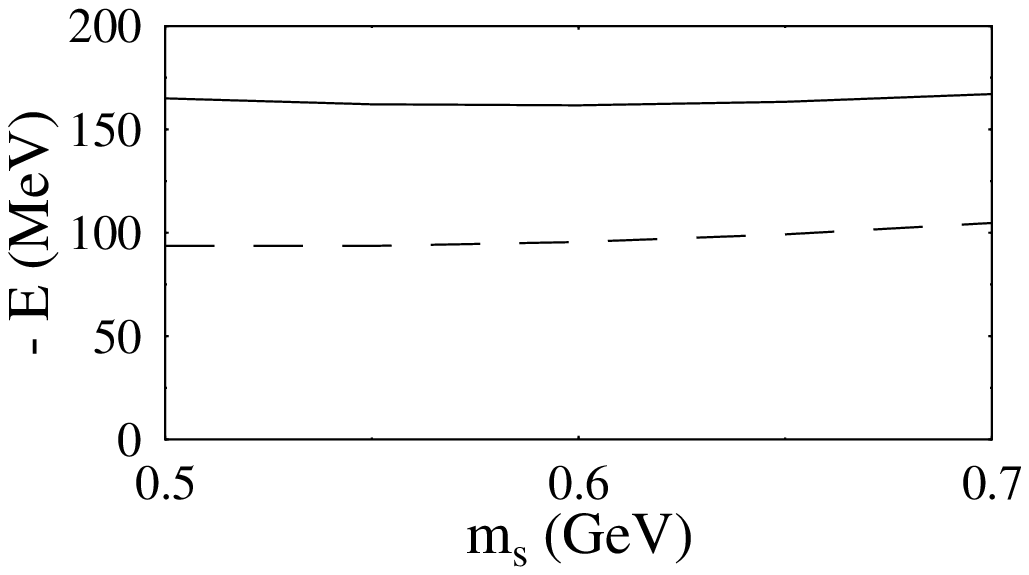}
	   }
	\caption{{\sf The mass $M$ and binding energy 
	$E=M-m_a-N_s m_s - N_v m_v$
	of the nucleon as a function of
	the scalar diquark mass, with $m_v=m_s +0.2\ GeV$ and 
	$m_a \equiv m_q =0.45\ GeV$
	. Dashed line: $g_s=2.17$, $g_v=1.64$, in accordance
	with \protect\cite{KLVW90,VW91}. Solid line: $g_s=2.0$, $g_v=1.51$.
	}}
	\label{fig:me}
\end{figure}
In the rest frame of the nucleon,
the scalar vertex functions depend 
on $\vert \vec{p} \vert$ and $p_0$. 
By taking appropriate traces of the BSE (\ref{eq:BSbig}), one obtains coupled
integral equations for the eight scalar vertex functions.

 The integrand of these kernels depend on $\vert \vec{p} \vert$, $p_0$ and 
on the angle $\theta$, defined by
$\vec{p} \cdot \vec{p}^\prime = \vert \vec{p} \vert \vert \vec{p}^\prime \vert \cos \theta$. 
This last dependence enters only through
the potential and the corresponding integral
can be done independently
of the form of the eight scalar vertex functions.
The integrands do not depend on
the azimuthal angle $\phi$. 
Poles in the propagators are avoided by performing a Wick-rotation: 
$p^0 \rightarrow i p^0$.
Now, one can start with eight arbitrarily chosen functions of
2 variables that are
defined on some $N_0 \otimes N_3$ grid. Applying the BSE 
repeatedly one finds a stable solution if $M$ has the correct value. In this
sense an eigenvalue problem for the relativistic energy is solved.

The parameters of the model are the scalar and vector diquark masses $m_s$ and
$m_v$, the scalar and vector coupling constants $g_s$ and $g_v$, the quark mass
$m_a$ and the Euclidean cutoff $\Lambda$. We take $m_a=0.45\ GeV$ for
the quark mass, suggested
by \cite{MM91} and 
$m_v-m_s=0.2\ GeV$ for the difference of the vector and scalar diquark masses,
which reproduces the $\Delta-N$ mass difference \cite{GRG75}. 
The cutoff is taken to be $\Lambda=0.9\ GeV$.

In figure (\ref{fig:me}) the mass $M$ and 
binding energy $E$ of the nucleon as a quark-diquark bound state
are shown as a function of $m_s$.
The binding energy is defined as 
$E=M-m_a-N_s m_s - N_v m_v$, where $N_s=\eta_a^s N_{qS} + \eta_b^s N_{dS}$ and
$N_v={3\over 2} ( \eta_a^v N_{qV} + \eta_b^v N_{dV} )$ are weighting factors
for the scalar and vector channels.
The nucleon mass is reached for a 
scalar diquark mass of about $0.6\ GeV$. This result however depends on
the parameters of the model. In order to constrain these parameters further, it
is necessary also to calculate other physical observables. 
Still it is interesting
to see that the binding energy hardly depends on the constituent masses and
is relatively low. Evidently in this picture most of the nucleon mass is
already given by the sum of quark and diquark quasiparticle masses.

If one puts the scalar coupling $g_s$ to zero by hand, it turns out that
the potential in the vector channel is repulsive.  
However, if the scalar coupling is turned on, the potential is attractive
in both the scalar and vector channels.

\begin{figure}
\centerline{	\epsfxsize=7cm
		\epsfbox{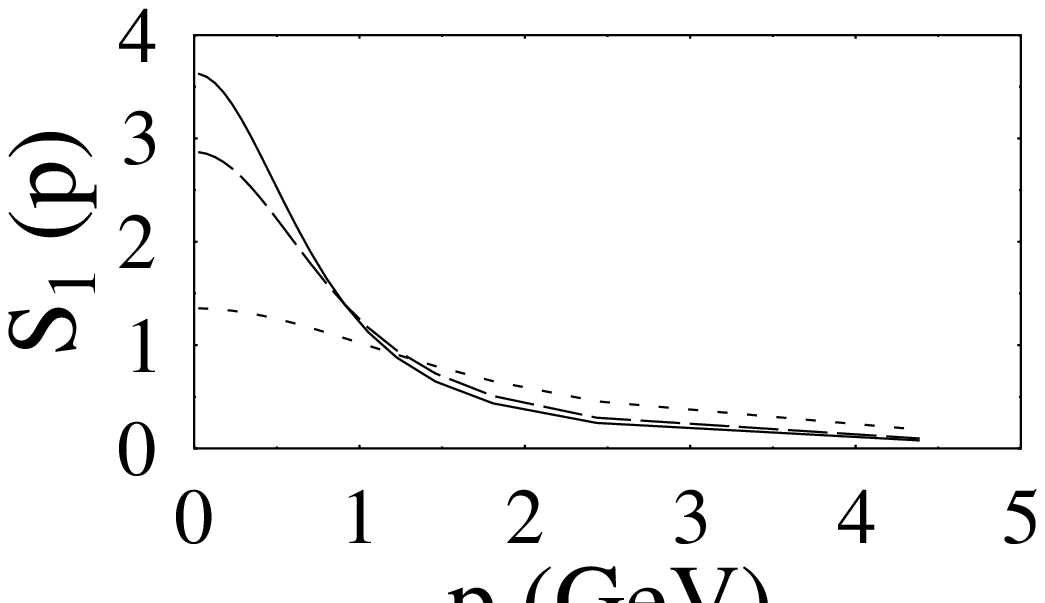}
		\epsfxsize=7cm
		\epsfbox{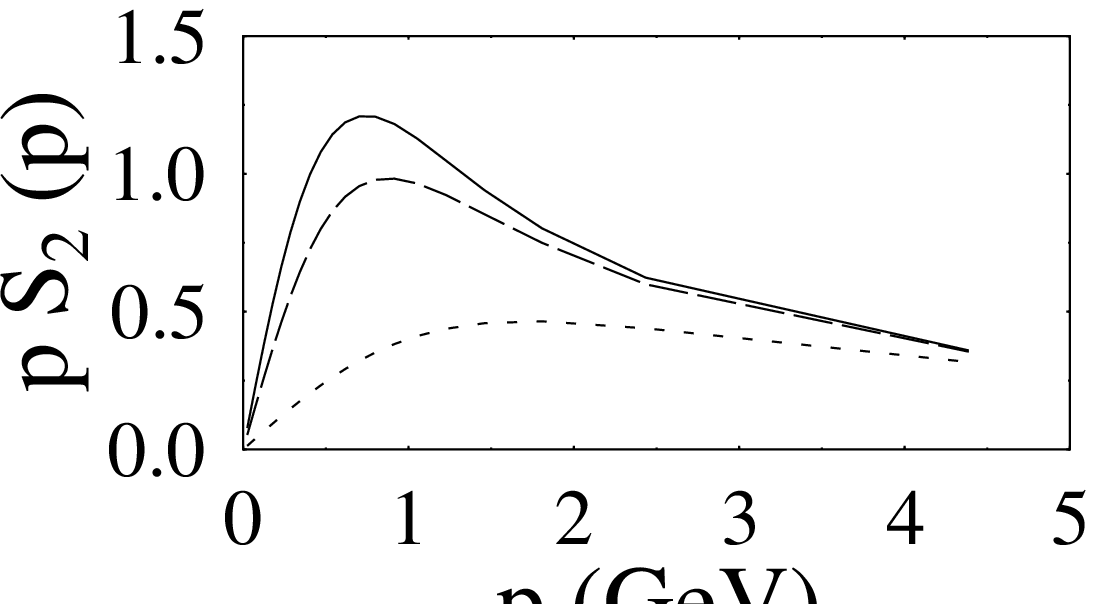}
	   }
\vspace{0.7cm}
\centerline{	\epsfxsize=7cm
		\epsfbox{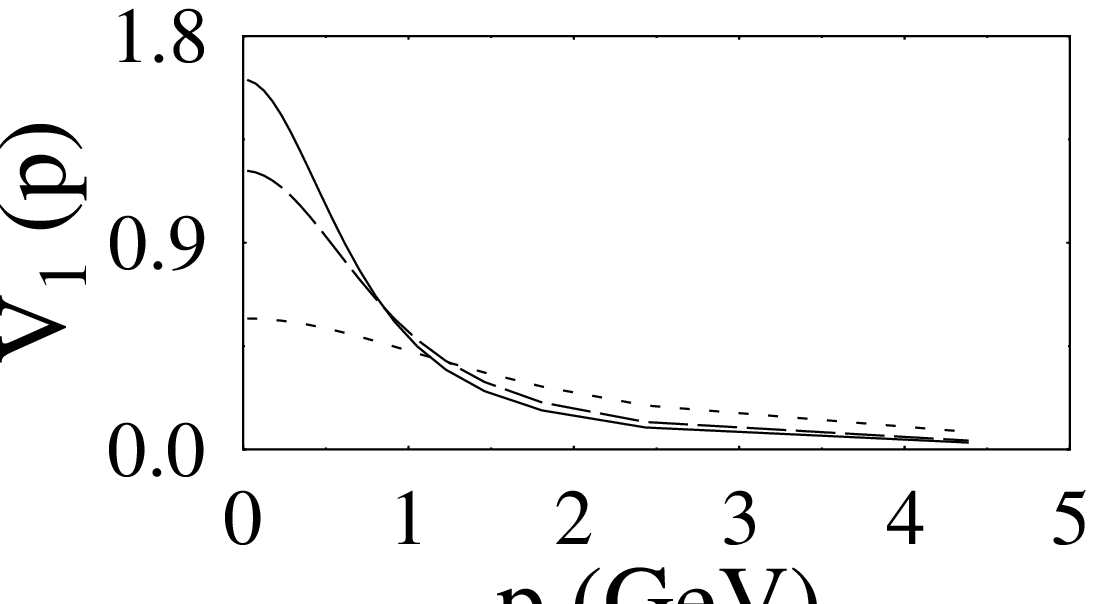}
		\epsfxsize=7cm
		\epsfbox{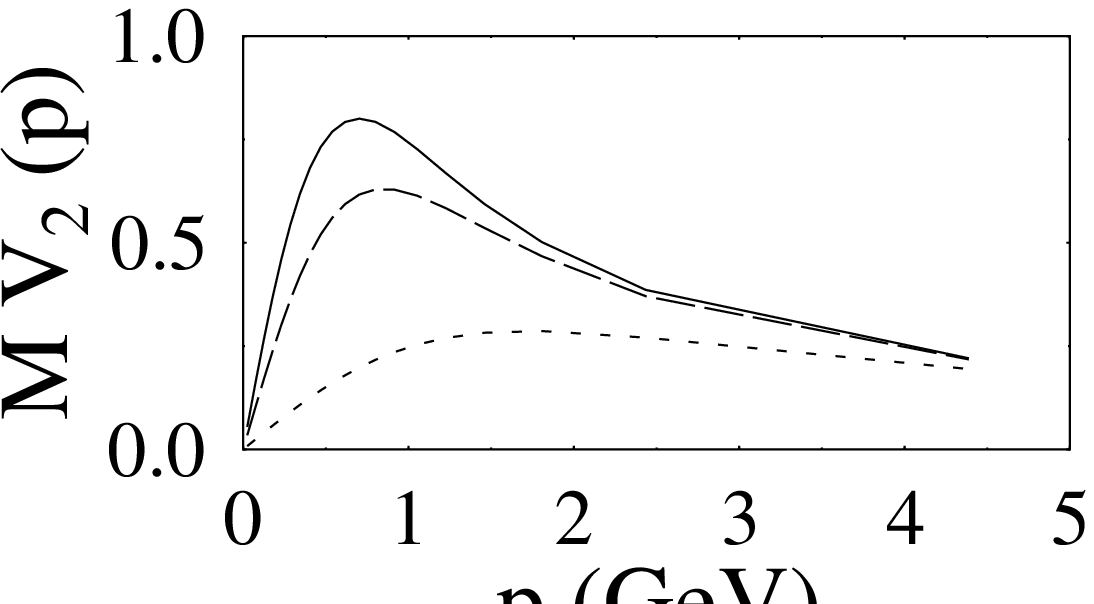}
	   }
\vspace{0.7cm}
\centerline{	\epsfxsize=7cm
		\epsfbox{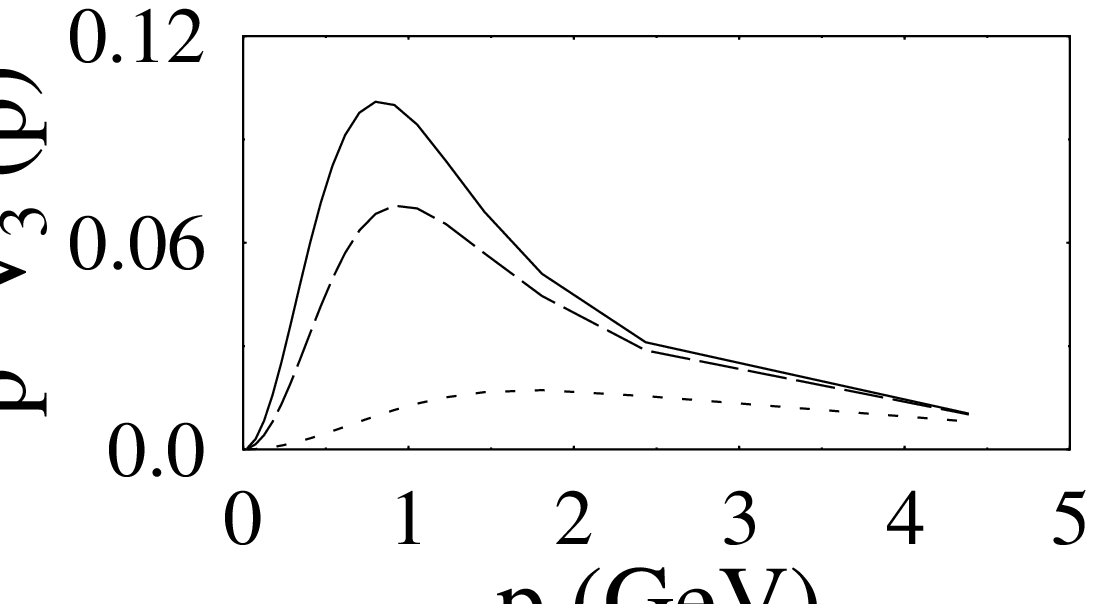}
		\epsfxsize=7cm
		\epsfbox{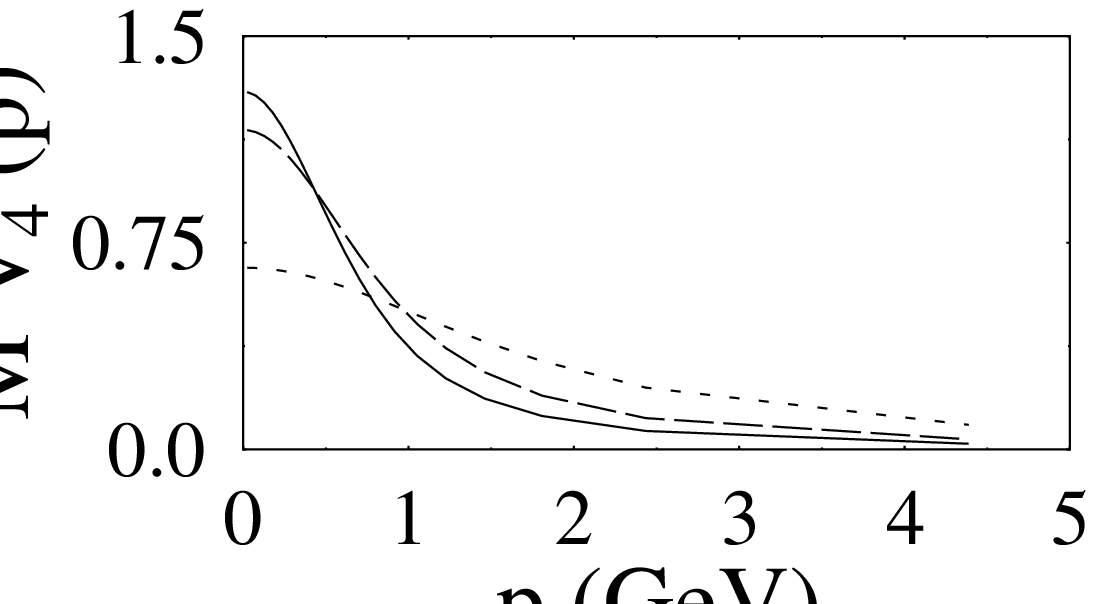}
	   }
\vspace{0.7cm}
\centerline{	\epsfxsize=7cm
		\epsfbox{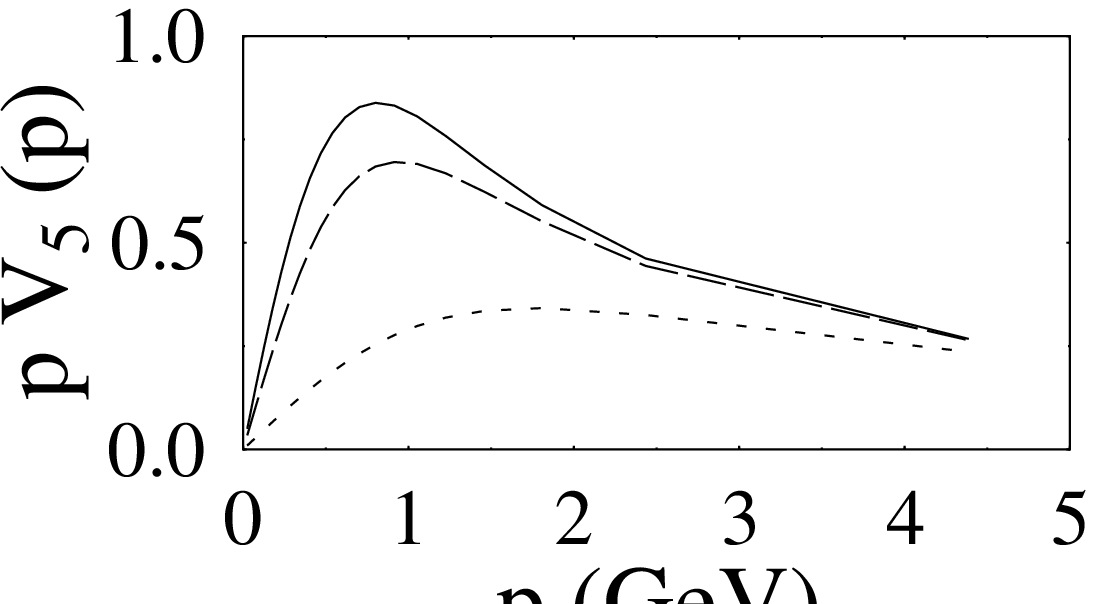}
		\epsfxsize=7cm
		\epsfbox{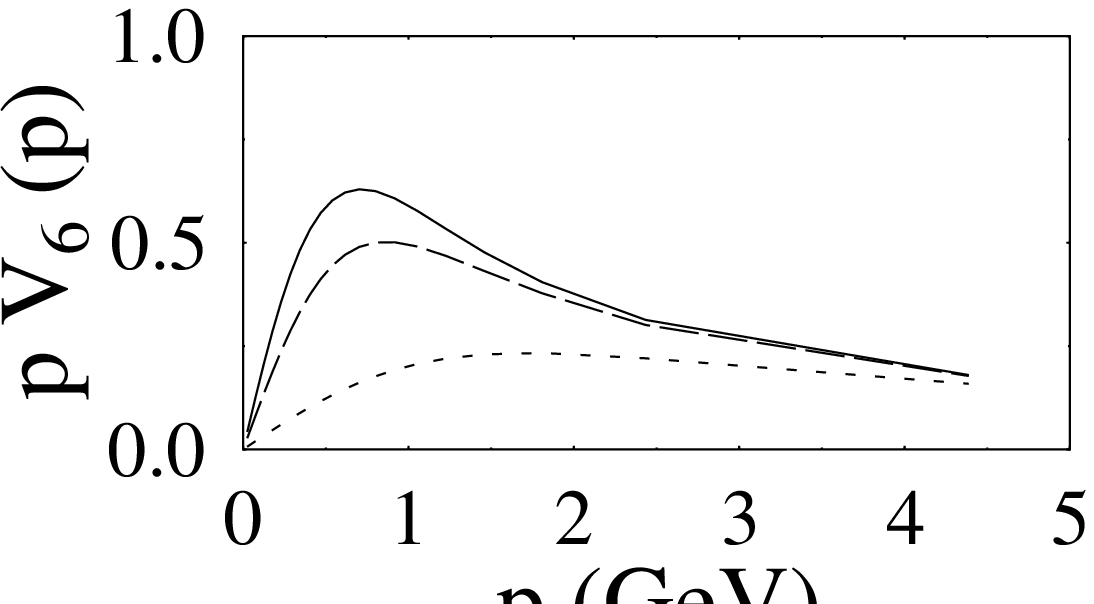}
	   }
\vspace{0.7cm}
	\caption{{\sf 
The Bethe-Salpeter
vertex functions as functions of the absolute value of the relative
 3-momentum $p$ for
three values of the quark-diquark
relative energy $p^0=0.05\ GeV$ (solid), $p^0=0.50\ GeV$ (dashed)
and $p^0=1.61\ GeV$ (dotted).
	}}
	\label{fig:vertex}
\end{figure}

The resulting Bethe-Salpeter vertex functions are
plotted in figure (\ref{fig:vertex}) 
as functions of the absolute value of the relative 3-momentum $p$ for
different values of the relative energy $p_0$. This second variable does
not have a non-relativistic analogue; the n.r. wave function is obtained
from the relativistic vertex by integrating over the relative energy.
The plotted functions include characteristic factors 
$\vert \vec{p} \vert ^l$ where $l=0,\ 1,\ 2$ correspond to s-, p- and
d-waves.
It is seen that the s-wave contributions
$S_1$, $V_1$ and $V_4$ dominate, but the p-wave contributions $S_2$,
$V_2$, $V_5$ and $V_6$ are also important. The d-wave term $V_3$ can be
neglected. In the static limit, only the s-waves survive. 

The values of the norm in the various channels are: 
\be
	N_{qS}=0.80,\
	N_{dS}=0.76,\
	N_{qV}=0.17,\
	N_{dV}=0.14.
\ee
One observes that the conditions $Q_p=1$ and $Q_n=0$ for the proton
and neutron charge (see eqs.(\ref{eq:Qp}, \ref{eq:Qn})) are very well
realized.

It is interesting to see that the charges carried by the S and V diquark
channels in the proton, 0.79 and 0.20 respectively, differ substantially
from the values obtained in the non-relativistic 
$SU(2)_{spin}\ \otimes\ SU(2)_{isospin}$ limit where these charges are ${1\over 2}$
for each channel.
This result indicates that the scalar
diquark component of the quark-diquark-nucleon vertex is more important
than the vector diquark component.

I would like to thank Wolfram Weise for fruitful discussions and careful
reading of the manuscript.

\bibliography{verderlese}
\end{document}